\documentclass[12pt,aps,nofootinbib,preprint,superscriptaddress]{revtex4}
\usepackage{amssymb}
\usepackage{amsmath}
\usepackage{amsfonts}
\usepackage{axodraw}

\newcommand{\eps}{\varepsilon}
\newcommand{\Refs}{Refs.}
\newcommand{\Ref}{Ref.}

\newcommand{\eq}{Eq.}
\newcommand{\eqs}{Eqs.}
\newcommand{\fig}{Fig.}
\newcommand{\figs}{Figs.}

\newcommand{\ie}{\emph{i.e.}}
\newcommand{\eg}{\emph{e.g.}}

\newcommand{\dv}{\partial\hspace{-7pt}\slash}
\newcommand{\be}{\begin{equation}}
\newcommand{\ee}{\end{equation}}
\newcommand{\bea}{\begin{eqnarray}}
\newcommand{\eea}{\end{eqnarray}}

\allowdisplaybreaks

\begin{document}

\title{Loop bounds on non-standard neutrino interactions}

\author{Carla Biggio}
\email[]{biggio@mppmu.mpg.de}
\affiliation{Max-Planck-Institut f\"ur Physik (Werner-Heisenberg-Institut),
F\"ohringer Ring 6, 80805 M\"unchen, Germany}
\affiliation{Dipartimento di Fisica, Universit\`a di Genova, 
via Dodecaneso 33, 16146 Genova, Italy}
\author{Mattias Blennow}
\email[]{blennow@mppmu.mpg.de}
\affiliation{Max-Planck-Institut f\"ur Physik (Werner-Heisenberg-Institut),
F\"ohringer Ring 6, 80805 M\"unchen, Germany}
\author{Enrique Fernandez-Martinez}
\email[]{enfmarti@mppmu.mpg.de}
\affiliation{Max-Planck-Institut f\"ur Physik (Werner-Heisenberg-Institut),
F\"ohringer Ring 6, 80805 M\"unchen, Germany}

\begin{abstract}
We reconsider the bounds on non-standard neutrino interactions with
matter which can be derived by constraining the four-charged-lepton
operators induced at the loop level. We find that these bounds are
model dependent.  Naturalness arguments can lead to much stronger
constraints than those presented in previous studies, while no
completely model-independent bounds can be derived. We will illustrate
how large loop-contributions to four-charged-lepton operators are
induced within a particular model that realizes gauge invariant
non-standard interactions and discuss conditions to avoid these
bounds.  These considerations mainly affect the $\mathcal O(10^{-4})$
constraint on the non-standard coupling strength $\eps_{e\mu}$, which is lost. 
The only model-independent constraints that can be derived are $\mathcal O(10^{-1})$.
However, significant cancellations are required in order to saturate this bound.
\end{abstract}

\pacs{}

\preprint{MPP-2008-127}
\preprint{GEF-TH-2-09}

\maketitle


\section{Introduction}

Neutrino non-standard interactions (NSI) were originally proposed~\cite{Wolfenstein:1977ue} as a mechanism to produce neutrino
flavour conversion in matter and considered as a
possible explanation for the solar and atmospheric neutrino deficits~\cite{Wolfenstein:1977ue,Mikheev:1986gs,Roulet:1991sm,Guzzo:1991hi,Brooijmans:1998py,GonzalezGarcia:1998hj,Bergmann:2000gp,Guzzo:2000kx,Guzzo:2001mi}. Although
it is now clear that such NSI cannot fully account for the observed
neutrino flavour conversion, the increasing sensitivity of neutrino
oscillation experiments to sub-leading effects has triggered a new
interest in them and their interference with neutrino oscillations at
present (\eg, K2K, MINOS, OPERA~\cite{Grossman:1995wx,Ota:2002na,Friedland:2005vy,Kitazawa:2006iq,Friedland:2006pi,Blennow:2007pu,EstebanPretel:2008qi,Blennow:2008ym}) and future (\eg,
SuperBeams, $\beta$Beams or Neutrino
Factories~\cite{GonzalezGarcia:2001mp,Gago:2001xg,Huber:2001zw,Ota:2001pw,Campanelli:2002cc,Blennow:2005qj,Kopp:2007mi,Kopp:2007ne,Ribeiro:2007ud,
Bandyopadhyay:2007kx,Ribeiro:2007jq,Kopp:2008ds,Malinsky:2008qn}) facilities. In
particular, the determination of the leptonic mixing angle
$\theta_{13}$ could be severely affected by degeneracies with the
non-standard parameters \cite{Huber:2001de,Huber:2002bi,Ohlsson:2008gx}.

Neutrino NSI can be described by effective four-fermion operators
of the form
\begin{equation}
\label{def}
\mathcal L_{\rm NSI} =
- 2\sqrt{2}G_F\eps^{ff'L,R}_{\alpha\beta}
(\overline{\nu_{\alpha}} \gamma^\mu \nu_{\beta})
(\overline{f_{L,R}}\gamma_\mu f'_{L,R})\, ,
\end{equation}
where $f$ and $f'$ are charged fermions with the same quantum numbers
and $L,R$ represent the chirality. For the rest of the paper we will denote with $\nu$ the left-handed neutrinos, $\ell$ the left-handed
charged leptons, $L$ the left-handed lepton doublets and $E$ the
right-handed charged leptons. For NSI of neutrinos with normal
matter, $f = f'$ can be either an electron, an up-quark or a
down-quark. While strong experimental bounds are present on the
corresponding four-charged-fermion interactions,
$(\overline{\ell_{\alpha}} \gamma^\mu
\ell_{\beta})(\overline{f_{L,R}}\gamma_\mu f'_{L,R})$, neutrino NSI
are much less constrained. Constraints on them have been derived in
\Ref~\cite{Davidson:2003ha,Barranco:2005ps,Barranco:2007ej} making use of both tree level and one-loop
processes. Here we reconsider the one-loop bounds, focusing on the
necessity of implementing the neutrino NSI in a gauge invariant way in
order to obtain a gauge independent result.

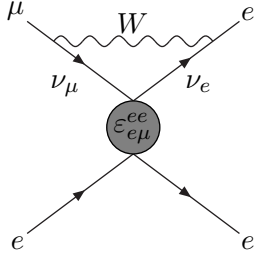
\begin{figure}
  \begin{center}
     \begin{picture}(100,100)(0,0)
       \ArrowLine(10,10)(50,40)
       \Text(9,9)[tr]{$e$}
       \ArrowLine(50,40)(90,10)
       \Text(91,9)[tl]{$e$}
       \ArrowLine(10,90)(50,60)
       \Text(9,91)[br]{$\mu$}
       \Text(30,70)[tr]{$\nu_\mu$}
       \ArrowLine(50,60)(90,90)
       \Text(91,91)[bl]{$e$}
       \Text(70,70)[tl]{$\nu_e$}
       \GCirc(50,50){10}{.5}
       \Text(50,50)[c]{$\eps^{ee}_{e\mu}$}
       \Photon(20,82.5)(80,82.5){2}{5.5}
       \Text(50,86)[b]{$W$}
     \end{picture}
    \caption{One-loop contribution to the four-charged-fermion vertex
      arising from the operator of \eq~(\ref{def}) via $W$
      exchange.}\label{fig:W}
  \end{center}
\end{figure}

The simple promotion of the neutrino fields in \eq~(\ref{def}) to
lepton doublets in order to construct a gauge invariant operator would
imply that the strong bounds stemming from flavour violating
four-charged-fermion processes will apply to neutrino interactions as
well. In order to avoid this and allow large NSI, the simultaneous presence of tree level flavour violating
four-charged-fermion
interactions must be forbidden~\cite{Berezhiani:2001rs,Antusch:2008tz,Gavela:2008ra}. However,
even if this requirement is satisfied, these interactions can be
generated at one loop from the operator of \eq~(\ref{def}) via a $W$
exchange between the neutrino legs. In \Ref~\cite{Davidson:2003ha}
this has been exploited to set bounds on some $\eps$. Notably, an
$\mathcal O(10^{-4})$ bound on $\eps_{e \mu}^{ff}$ was derived through
loop contributions to the decay $\mu \to 3e$ and the $\mu -e$
conversion in nuclei. Consequently,
the non-standard coupling strength $\eps^{ff}_{e\mu}$ was
neglected in a large number of studies
(see for example
\Refs~\cite{Friedland:2004pp,Friedland:2005vy,Friedland:2006pi,Kitazawa:2006iq,Blennow:2007pu,EstebanPretel:2007yu,EstebanPretel:2008qi,Blennow:2008eb,Kopp:2008ds,Winter:2008eg,Meloni:2009ia}). However, as we will show, when gauge invariance is imposed,
only naturalness arguments can be invoked. These arguments can lead to
even stronger constraints on the NSI strength, but no completely
model-independent bounds can be derived. The relevant diagram
contributing to $\mu \to 3e$ is depicted in \fig~\ref{fig:W}. The
computation of this diagram using only the operator of \eq~(\ref{def})
with $f=f'=e$ renders a gauge dependent result due to the fact that
the operator itself is not gauge invariant. Thus, the necessity of
considering a gauge invariant formulation of NSI is manifest.  In the
following, we will consider gauge invariant realisations of NSI both
when the operator of \eq~(\ref{def}) is realised from a dimension-six
($d=6$) and from a $d=8$ operator. We will discuss here the NSI with charged leptons, \ie, $f = f' =
\ell$. However, similar arguments as those presented here are
applicable in the case of neutrino NSI with quarks.
While gauge invariance has to be carefully taken care of in the cases mentioned above, one-loop processes like the 
$W$ and $Z$ decays (such as the ones depicted in \figs~5-7 of \Ref~\cite{Davidson:2003ha}) do not contain gauge boson propagators in the loops and the bounds derived from them still apply.


\section{$\boldsymbol{d=6}$ Realisations}

\subsection{Flavour Antisymmetric Operator}

There is only one gauge invariant $d=6$ effective operator that can
induce the effective interaction of \eq~(\ref{def}) in a direct way
while avoiding the generation of four-charged-fermion operators. This
operator is of the form:
\begin{equation}
\mathcal O^a_6 = 
(\bar L_\gamma i \tau_2 L^c_\alpha)(\overline{L^c_\beta} i \tau_2 L_\delta)\, ,
\end{equation}
where $L^c = C \bar L^T$ and $C$ is the charge conjugation operator.
It can be generated upon integrating out a heavy charged scalar
singlet with lepton number violating couplings to the lepton doublets,
as in \Ref~\cite{Zee:1980ai,Zee:1985rj,Bilenky:1993bt,Barbier:2004ez}.
This operator induces NSI only for leptons and with a very
characteristic flavour structure\footnote{NSI with right-handed
fields avoiding four-charged-fermion interactions can be realised through higher-dimensional operators as we will discuss in the next section.}:
\begin{equation}
2\mathcal O^a_6 = 
(\bar{\ell}_{\alpha} \gamma^\mu \ell_{\beta})(\bar{\nu}_\gamma \gamma_\mu \nu_{\delta})+
(\bar{\ell}_{\gamma} \gamma^\mu \ell_{\delta})(\bar{\nu}_\alpha \gamma_\mu \nu_{\beta})-
(\bar{\ell}_{\alpha} \gamma^\mu \ell_{\delta})(\bar{\nu}_\gamma \gamma_\mu \nu_{\beta})-
(\bar{\ell}_{\gamma} \gamma^\mu \ell_{\beta})(\bar{\nu}_\alpha \gamma_\mu \nu_{\delta})\, .
\end{equation}
This structure implies the antisymmetry relations
\begin{equation}
\label{antisymm}
\eps^{\alpha\beta}_{\gamma\delta} = 
-\eps^{\gamma\beta}_{\alpha\delta} = 
-\eps^{\alpha\delta}_{\gamma\beta} =
\eps^{\gamma\delta}_{\alpha\beta}\, .
\end{equation}
Thus, for each process of the type $\ell \to 3\ell^\prime$ (\ie,~a
decay of a heavy lepton into three lighter leptons), there will be
four contributing diagrams at one loop. For example,
the process $\tau^- \to \mu^-\mu^+ e^-$ will receive contributions
from $\eps^{\mu\tau}_{e \mu}$, $\eps^{e \mu}_{\mu \tau}$,
$\eps^{\mu\mu}_{e\tau}$ and $\eps^{e\tau}_{\mu\mu}$. In the
approximation of massless fermions, using the antisymmetry relations
of \eq~(\ref{antisymm}), the four contributions exactly cancel. Thus,
any loop contribution from $d = 6$ operators to the
four-charged-fermion interactions will be suppressed at least by a
factor $\mathcal O(m_{\ell}^2/M_W^2)$, where $m_\ell$ is the mass of
the heaviest lepton involved in the process. As a consequence, the
bound on the corresponding $\eps$ will be increased by the inverse of
this factor.  However, the antisymmetry relations of
\eq~(\ref{antisymm}) also imply that NSI with electrons that change
the neutrino flavour are related to charged lepton flavour changing
interactions, and thus stringent constraints can be derived in this
case (see \cite{Cuypers:1996ia,Antusch:2008tz}).  Note that $\mu \to
3e$ is forbidden for $d=6$ operators, since there are no $\eps$ with
the given symmetries which can contribute to this process.

\subsection{NSI from Non-Unitarity}

The second possibility to generate neutrino NSI
avoiding four-charged-fermion interactions is in an indirect way via
the dimension six operator
\begin{eqnarray}\label{Eq:Dim6Kin}
\mathcal O^{\rm kin}_6 = 
- (\bar L_\alpha \tilde{H}) \,i\dv\, (\tilde{H}^\dagger L_\beta)\, ,
\end{eqnarray} 
where $H$ is the Higgs doublet (we choose the hypercharge of $H$ to be
$1/2$) and $\tilde H = i\tau_2 H^*$. This operator induces
non-canonical neutrino kinetic terms. After diagonalising and
normalising them, a non-unitary leptonic mixing matrix is produced
and, upon integrating out the $W$ and $Z$ bosons, neutrino NSI are
induced. The tree level generation of this operator, avoiding similar
contributions to charged leptons (see, \eg, \Ref~\cite{Abada:2007ux}),
involves the addition of Standard Model-singlet fermions (right-handed
neutrinos) which couple to the Higgs and lepton doublets via Yukawa
couplings like in the standard seesaw
model~\cite{Minkowski:1977sc,Mohapatra:1979ia,Yanagida:1979as,GellMann:1980vs}.

This second realisation of NSI at $d=6$ is also quite constrained due
to the effects of a non-unitary mixing matrix (see
\Refs~\cite{Tommasini:1995ii,Antusch:2006vwa,Antusch:2008tz}). The
bounds on the $\eps_{e\mu}$ element are particularly strong due to
the enhancement of the $\mu \rightarrow e \gamma$ process if the GIM
mechanism is not realised given the non-unitarity of the mixing
matrix. In \Ref~\cite{Antusch:2008tz}, a bound of $|\eps_{e \mu}| < 5.9
\times 10^{-5} |n_n/n_e -1|$ for $M_{N_R} > M_W$ or $|\eps_{e \mu}| <
9.1 \times 10^{-4} |n_n/n_e -1|$ for $M_{N_R} < M_W$ was computed,
where $n_{n}$ ($n_{e}$) is the neutron (electron) density in
matter. Notice that, since $n_n \simeq n_e$, this means an additional
suppression of $\mathcal O(10^{-2})$. These bounds are already
much stronger than the loop bounds discussed here and thus we will not
consider this possibility further.


\section{$\boldsymbol{d=8}$ Realisations}

Dimension eight realisations of the NSI offer more freedom. Gauge
invariant $d=8$ operators can be generated by adding two Higgs
doublets to the four-lepton operators. The vev of the Higgs field can
then be exploited to break the $SU(2)$ symmetry between the charged
leptons and the neutrinos and can thus induce neutrino
NSI avoiding their charged-lepton counterparts. A basis for operators
involving two Higgs doublets, two left-handed lepton doublets and two
fermions, either left or right-handed (for the matter components), is
given in \Refs~\cite{Berezhiani:2001rs,Gavela:2008ra}:
\begin{eqnarray}
  \mathcal{O}_{LLH}^{{\bf 111}}
  &=&
  (\bar{L}_{\beta} \gamma^{\rho} L_{\alpha})
  (\bar{L}_{\delta} \gamma_{\rho} L_{\gamma})
  (H^{\dagger} H)\, , \label{equ:br3} \\
  \mathcal{O}_{LLH}^{{\bf 331}}
  &=&
  (\bar{L}_{\beta} \gamma^{\rho} \vec{\tau} L_{\alpha})
  (\bar{L}_{\delta} \gamma_{\rho} \vec{\tau} L_{\gamma})
  (H^{\dagger} H)\, , \label{equ:br4} \\
  \mathcal{O}_{LLH}^{{\bf 133}}
  &=&
  (\bar{L}_{\beta} \gamma^{\rho} L_{\alpha})
  (\bar{L}_{\delta} \gamma_{\rho} \vec{\tau} L_{\gamma})
  (H^{\dagger} \vec{\tau} H)\, , \label{equ:br5} \\
 \mathcal{O}_{LLH}^{{\bf 313}}
 &=&
  (\bar{L}_{\beta} \gamma^{\rho} \vec{\tau} L_{\alpha})
  (\bar{L}_{\delta} \gamma_{\rho}  L_{\gamma})
  (H^{\dagger} \vec{\tau} H)\, , \label{equ:br6} \\
  \mathcal{O}_{LLH}^{{\bf 333}}
  &=&
  (-{\rm i}\epsilon^{abc})
  (\bar{L}_{\beta} \gamma^{\rho} \tau^{a} L_{\alpha})
  (\bar{L}_{\delta} \gamma_{\rho} \tau^{b} L_{\gamma})
  (H^{\dagger} \tau^{c} H)\, , \label{equ:br7} \\
 \mathcal{O}_{LEH}^{\bf 111}
 &=&
 (\bar{L}_{\beta} \gamma^{\rho}  L_{\alpha})
 (\bar{E}_{\delta} \gamma_{\rho}  E_{\gamma})
 \left(H^{\dagger} H\right)\, , \label{equ:br1} \\
 \mathcal{O}_{LEH}^{\bf 313}
 &=&
 (\bar{L}_{\beta} \gamma^{\rho} \vec{\tau}  L_{\alpha})
 (\bar{E}_{\delta} \gamma_{\rho}  E_{\gamma})
 \left(H^{\dagger} \vec{\tau} H \right)\, . \label{equ:br2} \, 
\end{eqnarray}
In addition to the two left-handed lepton doublets, the two last
operators contain two right-handed charged leptons and the first five
two additional left-handed lepton doublets. The generalisation to
operators involving interactions with quarks is straightforward
replacing these fields by their quark counterparts. Generically,
after electro-weak symmetry breaking (EWSB), these operators generate
both neutrino NSI and non-standard four-charged-fermion interactions
at tree level. In order to avoid the latter, the following conditions
have to be met~\cite{Gavela:2008ra}\footnote{Notice that our
relations differ from those in \Ref~\cite{Gavela:2008ra} due to different
conventions for the Higgs hyper-charge.}:
\be
 \mathcal{C}_{LEH}^{{\bf 111}} = -\mathcal{C}_{LEH}^{\bf 313} \, , \quad 
\mathcal{C}_{LLH}^{{\bf 111}}
 +
 \mathcal{C}_{LLH}^{{\bf 331}}
 +
 \mathcal{C}_{LLH}^{{\bf 133}}
 +
 \mathcal{C}_{LLH}^{{\bf 313}} = 0 \, , \quad
 \mathcal{C}_{LLH}^{{\bf 333}} \, \, \text{arbitr.} \, .
\label{equ:canceleight}
\ee 
The second condition involves four operators of the basis and we can
thus choose three independent combinations satisfying it in order to
form a basis for the $d=8$ operators that induce neutrino NSI avoiding
four-charged-lepton interactions:
\begin{eqnarray}
\label{A}
\mathcal{A}&=&
\frac 14(\mathcal{O}_{LLH}^{\bf 111}-\mathcal{O}_{LLH}^{\bf 331}) = 
(\overline{L^c_\alpha} i \tau_2 L_\gamma)(\bar L_\delta i \tau_2 L^{c}_\beta)(H^{\dagger} H)\, , \\
\label{B}
\mathcal{B}&=&
(\mathcal{O}_{LLH}^{\bf 111}-\mathcal{O}_{LLH}^{\bf 133})=2
(\bar{L}_{\beta}\gamma_{\rho} L_{\alpha})(\bar{L}_{\delta} \tilde{H}) \gamma^{\rho}
(\tilde{H}^\dagger L_{\gamma})\, , \\
\label{C}
\mathcal{C}&=&
(\mathcal{O}_{LLH}^{\bf 111}-\mathcal{O}_{LLH}^{\bf 313})=2
(\bar{L}_{\beta} \tilde{H}) \gamma^{\rho} (\tilde{H}^\dagger L_{\alpha})
 (\bar{L}_{\delta} \gamma_{\rho}  L_{\gamma})\, , \\
\label{D}
\mathcal{D}&=&
\mathcal{O}_{LLH}^{\bf 333}= 
2(\bar{L}_{\beta}\gamma_{\rho} L_{\gamma})(\bar{L}_{\delta} \tilde{H}) 
\gamma^{\rho}(\tilde{H}^\dagger L_{\alpha})
-2(\bar{L}_{\beta} \tilde{H}) \gamma^{\rho} (\tilde{H}^\dagger L_{\gamma})
 (\bar{L}_{\delta} \gamma_{\rho}  L_{\alpha})
 \, ,\\
\label{E}
\mathcal{E}&=&
(\mathcal{O}_{LEH}^{\bf 111}-\mathcal{O}_{LEH}^{\bf 313}) = 2
(\bar{L}_{\beta}\tilde{H}) \gamma^{\rho} (\tilde{H}^\dagger L_{\alpha})
(\bar{E}_{\delta} \gamma_{\rho} E_{\gamma})\, .
\end{eqnarray} 
All of these operators have to be divided by $M^4$, where $M$ is the
new physics scale from which the NSI originate, and multiplied by a
flavour-dependent coefficient: $a^{\alpha\beta\gamma\delta}$
multiplies $\mathcal{A}$, $b^{\alpha\beta\gamma\delta}$ multiplies
$\mathcal{B}$ and so on.  Note that, apart from $\mathcal{E}$, which
is independent since it is the only one containing right-handed
fields, the other operators are all related. Indeed $\mathcal{C}$ is
obtained from $\mathcal{B}$ simply by rearranging the flavour indexes,
$\mathcal{D}$ is equivalent to $\mathcal{B}-\mathcal{C}$ after a Fierz
transformation and again a rearrangement of two indexes and also
$\mathcal{A}$ can be written in terms of combination of $\mathcal{B}$
operators with different flavour indexes. Thus, the only new structure
that $d=8$ operators offer to select neutrino NSI avoiding
four-charged-fermion operators is the combination $(\bar{L} \tilde{H})
\gamma^{\rho} (\tilde{H}^\dagger L)$ present in
\eqs~(\ref{B})-(\ref{E}). 

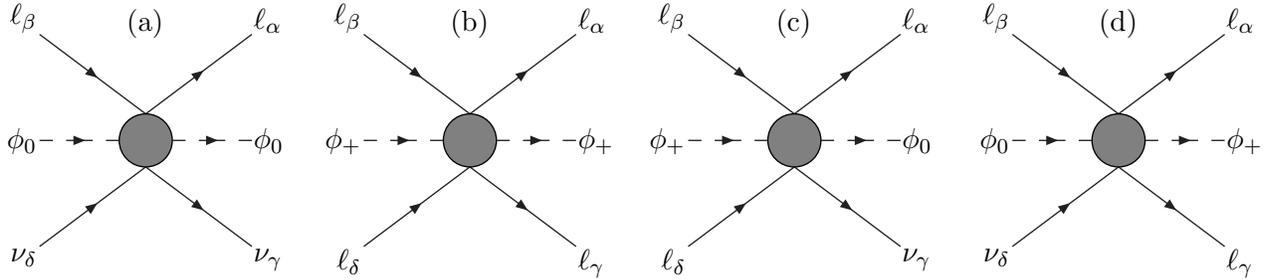
\begin{figure}
  \begin{center}
    \begin{picture}(100,100)(0,0)
      \Text(50,100)[t]{(a)}
      \ArrowLine(10,10)(50,40)
      \Text(9,9)[tr]{$\nu_\delta$}
      \ArrowLine(50,40)(90,10)
      \Text(91,9)[tl]{$\nu_\gamma$}
      \ArrowLine(10,90)(50,60)
      \Text(9,91)[br]{$\ell_\beta$}
      \ArrowLine(50,60)(90,90)
      \Text(91,91)[bl]{$\ell_\alpha$}
      \GCirc(50,50){10}{.5}
      \DashArrowLine(10,50)(40,50){4}
      \Text(9,50)[r]{$\phi_0$}
      \DashArrowLine(60,50)(90,50){4}
      \Text(91,50)[l]{$\phi_0$}
    \end{picture}
    \hspace{\stretch{1}}
    \begin{picture}(100,100)(0,0)
      \Text(50,100)[t]{(b)}
      \ArrowLine(10,10)(50,40)
      \Text(9,9)[tr]{$\ell_\delta$}
      \ArrowLine(50,40)(90,10)
      \Text(91,9)[tl]{$\ell_\gamma$}
      \ArrowLine(10,90)(50,60)
      \Text(9,91)[br]{$\ell_\beta$}
      \ArrowLine(50,60)(90,90)
      \Text(91,91)[bl]{$\ell_\alpha$}
      \GCirc(50,50){10}{.5}
      \DashArrowLine(10,50)(40,50){4}
      \Text(9,50)[r]{$\phi_+$}
      \DashArrowLine(60,50)(90,50){4}
      \Text(91,50)[l]{$\phi_+$}
    \end{picture}
    \hspace{\stretch{1}}
    \begin{picture}(100,100)(0,0)
      \Text(50,100)[t]{(c)}
      \ArrowLine(10,10)(50,40)
      \Text(9,9)[tr]{$\ell_\delta$}
      \ArrowLine(50,40)(90,10)
      \Text(91,9)[tl]{$\nu_\gamma$}
      \ArrowLine(10,90)(50,60)
      \Text(9,91)[br]{$\ell_\beta$}
      \ArrowLine(50,60)(90,90)
      \Text(91,91)[bl]{$\ell_\alpha$}
      \GCirc(50,50){10}{.5}
      \DashArrowLine(10,50)(40,50){4}
      \Text(9,50)[r]{$\phi_+$}
      \DashArrowLine(60,50)(90,50){4}
      \Text(91,50)[l]{$\phi_0$}
    \end{picture}
    \hspace{\stretch{1}}
    \begin{picture}(100,100)(0,0)
      \Text(50,100)[t]{(d)}
      \ArrowLine(10,10)(50,40)
      \Text(9,9)[tr]{$\nu_\delta$}
      \ArrowLine(50,40)(90,10)
      \Text(91,9)[tl]{$\ell_\gamma$}
      \ArrowLine(10,90)(50,60)
      \Text(9,91)[br]{$\ell_\beta$}
      \ArrowLine(50,60)(90,90)
      \Text(91,91)[bl]{$\ell_\alpha$}
      \GCirc(50,50){10}{.5}
      \DashArrowLine(10,50)(40,50){4}
      \Text(9,50)[r]{$\phi_0$}
      \DashArrowLine(60,50)(90,50){4}
      \Text(91,50)[l]{$\phi_+$}
    \end{picture}
    \caption{The effective interactions induced by $\mathcal B$.}
\label{fig:treelevel}
  \end{center}
\end{figure}

We will therefore compute the loop of \fig~\ref{fig:W} for this new
structure. From the combination
$(\bar{L}\gamma_{\rho}L)(\bar{L} \tilde{H})
\gamma^{\rho} (\tilde{H}^\dagger L)$, ``opening'' the operators in
components, it is easy to check which operators are generated together
with the one in \eq~(\ref{def}). Taking all of them into account, a
gauge invariant computation can be performed.  In this case, the
generated $\eps$ are completely independent and the effective
interactions shown in \fig~\ref{fig:treelevel} are all generated with
the same strength.
Out of these diagrams, diagram (a) will result in the effective
neutrino NSI when the Higgs acquires a
vev. There are now several one-loop contributions to the four
charged-lepton vertex. In diagram (a), we can connect the neutrino
lines with either a $W$ or a $\phi_+$, resulting in a conversion of
the neutrinos into charged leptons. For diagrams (c) and (d), we can
connect the charged Goldstones to the neutrino lines, and finally, for
diagram (b) we can close the $\phi_+$-loop. It is now possible to
check the gauge independence of the result explicitly. We have done
this by performing the computation in the $R_\xi$ gauge and splitting
the $W$ propagator in the unitary gauge part and the $\xi$-dependent part
and checked that the gauge dependence introduced by the $W$ and all
the diagrams with the Goldstones cancels, as it should. The remaining
gauge independent contribution is then the one from the $W$ exchange
with the propagator in the unitary gauge.

The most widespread way of estimating bounds from loop processes of an
unknown high energy theory through its effective description is
exploiting the logarithmic divergence, as in
\Ref~\cite{Davidson:2003ha}. Indeed the coefficient of this term and
that of the logarithmic running of the full theory are the same and
the mild scale dependence is just an $\mathcal{O}(1)$ correction. On
the other hand, the finite and quadratic contributions of the effective
theory are less reliable, since they depend on the matching with the
unknown full theory. Moreover, these contributions can be fine-tuned
away, while for the logarithmic running this can only be true at a
given scale\footnote{Unless the cancellation is implemented through
  an operator with exactly the same running.}.

We find that, neglecting lepton masses, the logarithmic divergences
coming from the two parts of the $W$-propagator exactly cancel at one 
loop.\footnote{Nevertheless, they might be present at the two loop level.} Then,
also in the case of this $d=8$ operator, only very weak bounds on
$\eps$ can be derived through the logarithmic divergence using the
decay of a heavy lepton into three lighter leptons. It should be noted
that, even if the logarithmic divergence is not present, a quadratic
divergence is. We will devote the next subsection to the physical
interpretation of this quadratic divergence.

We argue that the discussion presented here can be also applied to
higher-dimensional operators where only Higgs doublets are added. In
order to preserve gauge invariance in the NSI, it is necessary to
include the effects of the internal Goldstone loops. In order to avoid
specifying the underlying theory, we must therefore compute the loop
diagrams of the effective theory in the unitary gauge, where the
Goldstone propagators vanish.

\subsection{The quadratic divergence}

The computation of the loop with a $W$ exchange between the neutrino
legs of diagram (a) of \fig~\ref{fig:treelevel} in the unitary gauge
turns out to give a vanishing logarithmic contribution, while a
quadratic divergence is present. It is easy to check that a similar
diagram with a $Z$ exchange plus a diagram where the physical Higgses
of diagram (a) of \fig~\ref{fig:treelevel} are closed in a loop give
exactly the same quadratic divergence to the neutrino NSI operator.

It is simpler to understand the origin of this quadratic contribution
to both operators from the decomposition into singlet minus triplet of
\eqs~(\ref{B})-(\ref{E}):
\begin{equation}
(\bar{L} \tilde{H}) \gamma^{\rho} (\tilde{H}^\dagger L) = \frac{1}{2} [(\bar{L} \gamma^{\rho} L)(H^\dagger H) - (\bar{L} \gamma^{\rho} \vec{\tau} L)(H^\dagger \vec{\tau} H)]\, .
\end{equation}
If the Higgs legs are closed in a loop, the resulting tadpole will
contribute with a quadratic divergence only through the singlet
term. The corresponding divergence is not present for the triplet term
since it involves a trace of the $\tau$ matrices, which vanishes. This
means that the quadratic divergence will be proportional to the
singlet combination $(\bar{L} \gamma^{\rho} L)$ and thus will be equal
for the neutrino NSI and the four-charged-fermion interaction. Since
the logarithmic contributions to the process vanish, we can now try to
estimate the constraints that this quadratic divergence implies. We
will consider these quadratic contributions for the operator of
\eq~(\ref{E}) involving the right-handed leptons $E$, but similar
discussions also apply to \eqs~(\ref{B})-(\ref{D}), as well as to interactions with quarks,
while the contributions cancel for \eq~(\ref{A}). However, as we will show
later, for the left-handed fields antisymmetries similar to the ones
discussed at $d=6$ can be considered in order to avoid the bounds.

At tree level, after EWSB, the operator $\mathcal{E}$ generates
neutrino NSI with strength
\begin{equation}
\label{epstree}
\eps^{\delta \gamma}_{\beta \alpha}=
-\frac{e^{\beta \alpha\delta \gamma}}{2} \frac{v^4}{M^4}\, .
\end{equation}
On the other hand, if we parametrise the four-charged-lepton interactions as
\begin{equation}
\label{defCF}
\mathcal L_{\rm CF} = - 2\sqrt{2}G_F\eps^{\beta\alpha,\gamma\delta}_{CF,R}
(\overline{\ell_{\beta}}\gamma_\mu \ell_{\alpha})
(\overline{E_{\delta}} \gamma^\mu E_{\gamma})\, ,
\end{equation} 
then $\eps^{\beta\alpha,\delta\gamma}_{CF,R} = 0$.  However, the
contribution from the quadratic divergence arising at
one loop is equal for the neutrino and the four-charged-lepton
NSI. Adding this, we have:
\begin{eqnarray}
\label{eps1loop}
\eps_{\beta \alpha}^{\delta \gamma}&=&
-\frac{e^{\beta \alpha\delta \gamma}}{2} \frac{v^2}{M^2}
\left(\frac{v^2}{M^2}+\frac{\Lambda^2}{M^2}\frac{k}{8\pi^2}\right)\\
\label{epsCF1loop}
\eps^{\beta \alpha,\delta\gamma}_{CF,R} &=&
-\frac{e^{\beta \alpha\delta \gamma}}{2}\frac{v^2}{M^2}
\frac{\Lambda^2}{M^2}\frac{k}{8\pi^2}\, ,
\end{eqnarray}
where $k = \mathcal O(1)$ depends
on the UV completion of the full theory, and $\Lambda
\le M$ is the scale at which new physics appears and the effective
theory is no longer valid. Thus, it is clear that no model independent
bounds can be derived from this quadratic contribution, since
assumptions on the sizes of $\Lambda$ and $k$ have to be made. For
example, the high energy completion of the theory could be such that
$k=0$ due to some significant fine-tunings and then no bounds would
stem from this process. On the other hand, naturalness arguments can
be invoked to argue that, in absence of significant fine-tunings, $k=
\mathcal O(1)$ and $\Lambda$ should be at least as high as the
electroweak scale. With these assumptions, bounds of the same order as
the ones derived in \Ref~\cite{Davidson:2003ha} will be recovered. On the other hand, if no new
physics is present between $v$ and $M$, $\Lambda$ could be identified
with $M$ and very stringent constraints could be derived.

Below we will consider, as an example, the computation of the
quadratic contribution in a complete theory whose low energy effects
are described precisely by the operator of \eq~(\ref{B}), without the
need of fine-tuning in order to cancel similar operators contributing
to four-charged-fermion processes at tree level. In this example, $\Lambda=M$ and
$k=1/2$. Thus, even if only neutrino NSI are
induced at tree level, the loop contribution only has a suppression of
$8\pi^2 \sim \mathcal O(100)$, which could dominate the
$\mathcal O(v^2/M^2)$ tree level contribution (unless $M\lesssim$
1~TeV) and the four-charged-fermion operator would be induced with a
strength similar to that of the neutrino NSI. This would imply strong
bounds on the neutrino NSI and the main motivation for considering
$d=8$ operators would be lost.

We will now discuss how some antisymmetries between the
$\eps$ parameters could avoid these corrections, in a way similar to
the $d=6$ antisymmetric realisation. As for $d=6$, this is only
possible for the case in which the matter fermions are left-handed
leptons. To study this,
it is
more convenient to change the operator basis to a basis where these
symmetries are manifest:
\begin{eqnarray}
   \mathcal A &=& \frac{1}{2}
    (|\phi_0|^2 + |\phi_+|^2)[
    (\bar\nu_\beta \gamma^\rho\nu_\alpha)(\bar\ell_\delta \gamma_\rho \ell_\gamma)
    +(\bar\nu_\delta \gamma^\rho\nu_\gamma)(\bar\ell_\beta \gamma_\rho \ell_\alpha) \nonumber \\
    &&
    \phantom{2(|\phi_0|^2 + |\phi_+|^2)}
    -(\bar\nu_\beta \gamma^\rho\nu_\gamma)(\bar\ell_\delta \gamma_\rho \ell_\alpha)
    -(\bar\nu_\delta \gamma^\rho\nu_\alpha)(\bar\ell_\beta \gamma_\rho \ell_\gamma)
  ]\, , 
\label{A-simm}\\
\nonumber
   \mathcal S &=& \frac{\mathcal B +\mathcal C}{2} - \mathcal A = \\
  && \frac{1}{2}
    (|\phi_0|^2+|\phi_+|^2)[
    (\bar\nu_\beta \gamma^\rho\nu_\alpha)(\bar\ell_\delta \gamma_\rho \ell_\gamma)
    +(\bar\nu_\delta \gamma^\rho\nu_\gamma)(\bar\ell_\beta \gamma_\rho \ell_\alpha) \nonumber \\
    &&
    \phantom{2(|\phi_0|^2+|\phi_+|^2)}
    +(\bar\nu_\beta\gamma^\rho \nu_\gamma)(\bar\ell_\delta \gamma_\rho \ell_\alpha)
    +(\bar\nu_\delta\gamma^\rho \nu_\alpha)(\bar\ell_\beta \gamma_\rho \ell_\gamma)
  ] \nonumber \\
  && +2
\left[|\phi_0|^2 (\bar\nu_\beta\gamma^\rho \nu_\alpha)(\bar\nu_\delta \gamma_\rho \nu_\gamma)
  +|\phi_+|^2 (\bar\ell_\beta\gamma^\rho \ell_\alpha)(\bar\ell_\delta \gamma_\rho \ell_\gamma)\right] + \ldots\, , 
\label{S-simm}\\
\nonumber
\label{X-simm}
   \mathcal X &=& \frac{-\mathcal B +\mathcal C}{2} = \\
&&  
 (|\phi_+|^2-|\phi_0|^2)[
    (\bar\ell_\beta\gamma^\rho \ell_\alpha)(\bar\nu_\delta \gamma_\rho \nu_\gamma)
    - (\bar\ell_\delta \gamma^\rho\ell_\gamma)(\bar\nu_\beta \gamma_\rho \nu_\alpha)
  ] + \ldots\, ,
\end{eqnarray}
where $\ldots$ represents terms proportional to $\phi_0^*\phi_+$ or
$\phi_0 \phi_+^*$, which will not contribute to our one-loop
computations as long as we consider the leptons to be
massless.

In the Feynman gauge, the quadratic divergences are completely
determined by the loops of the Goldstones and Higgs fields. Indeed,
since the quadratic divergence is independent of the mass propagating
in the loop, the terms $|\phi_0|^2$ (via the Higgs and neutral
Goldstone) and $|\phi_+|^2$ (via the charged Goldstones) will give the
same contribution. It is thus evident, by looking at
\eqs~(\ref{A-simm})-(\ref{X-simm}) that the operators $\mathcal S$ and
$\mathcal A$ will contain the quadratic divergence, while $\mathcal X$
will not, since the two contributions cancel.

Also, a generic coupling $c^{\alpha\beta\gamma\delta}$ can be
decomposed as
\begin{equation}
 c^{\alpha\beta\gamma\delta} = s^{\alpha\beta\gamma\delta} 
+ a^{\alpha\beta\gamma\delta} + x^{\alpha\beta\gamma\delta}\, ,
\label{decomposition}
\end{equation}
where
\begin{eqnarray}
  s^{\alpha\beta\gamma\delta} = s^{\gamma\beta\alpha\delta} 
= s^{\alpha\delta\gamma\beta} = s^{\gamma\delta\alpha\beta}
  &=&
  \frac 14 (c^{\alpha\beta\gamma\delta} + c^{\gamma\beta\alpha\delta} 
          + c^{\alpha\delta\gamma\beta} + c^{\gamma\delta\alpha\beta})\, , \\
  x^{\alpha\beta\gamma\delta} = - x^{\gamma\delta\alpha\beta}
  &=&
  \frac 12 (c^{\alpha\beta\gamma\delta} - c^{\gamma\delta\alpha\beta})\, , \\
  a^{\alpha\beta\gamma\delta} = -a^{\gamma\beta\alpha\delta} 
= -a^{\alpha\delta\gamma\beta} = a^{\gamma\delta\alpha\beta}
  &=&
  \frac 14 (c^{\alpha\beta\gamma\delta} - c^{\gamma\beta\alpha\delta} 
          - c^{\alpha\delta\gamma\beta} + c^{\gamma\delta\alpha\beta})\, .
\end{eqnarray}
We see that $s$ has the same symmetry as $\mathcal S$, $a$ has the
symmetry of $\mathcal A$, and $x$ has the symmetry of $\mathcal
X$, and are actually the coefficients of $\mathcal S$, $\mathcal A$, and $\mathcal X$.

\begin{figure}[t!]
\vspace{-0.5cm}
\begin{center}
\begin{picture}(250,100)(0,0)
	\Text(27,77)[lb]{$L_\alpha$}
	\ArrowLine(0,100)(50,50)
	\Text(27,23)[lt]{$\tilde H^\dagger$}
	\DashArrowLine(0,0)(50,50){4}
	\Text(75,46)[t]{$N_R$}
	\Line(50,50)(100,50)
	\Vertex(50,50){2}
	\Text(97,90)[r]{$\overline{L_\gamma^c}$}
	\ArrowLine(100,100)(100,50)
	\Text(125,46)[t]{$S$}
	\DashArrowLine(150,50)(100,50){4}
	\Vertex(100,50){2}
	\Text(175,46)[t]{$N_R$}
	\Line(150,50)(200,50)
	\Text(153,90)[l]{$L_\delta^c$}
	\ArrowLine(150,50)(150,100)
	\Vertex(150,50){2}
	\Text(223,77)[br]{$\overline{L_\beta}$}
	\ArrowLine(200,50)(250,100)
	\Text(223,23)[tr]{$\tilde H$}
	\DashArrowLine(200,50)(250,0){4}
	\Vertex(200,50){2}
\end{picture}
\caption{An example of a full theory inducing neutrino NSI, but not four-charged-fermion interactions, at tree level. The NSI are generated through the exchange of right-handed neutrinos $N_R$ and a scalar doublet $S$.\label{fig:d8}}
\end{center}
\end{figure}
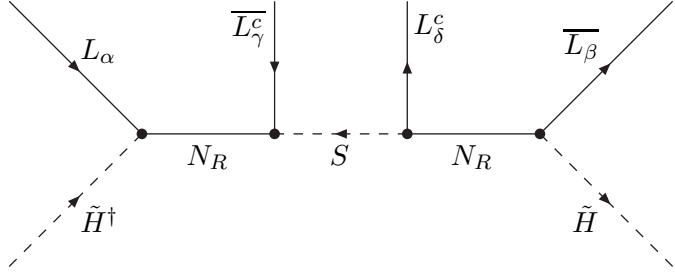

To illustrate how operators with different symmetries contribute in a
different way to the quadratic divergence, we will take the theory
from Section 5.1 of \Ref~\cite{Antusch:2008tz} as an example (see \fig~\ref{fig:d8}). 
In this theory, the NSI are realised at tree level by two Higgs doublets
selecting the neutrinos from two lepton doublets by taking a vev and the exchange of two right-handed neutrinos and a scalar doublet. For simplicity, we will here assume
that the scalar has the same mass $M$ as the right-handed
neutrinos. The effective tree level operator is now essentially given
by
\begin{equation}
  \label{eq:decomp}
\frac{2c^{\beta\alpha\delta\gamma}}{M^4}  
(\bar L_\beta \tilde H)\gamma^\rho (\tilde{H}^\dagger L_\alpha) 
(\bar L_\delta \gamma_\rho L_\gamma)
= \frac{1}{M^4} (\mathcal S + \mathcal A + \mathcal X)
\end{equation}
and thus contains all of the above mentioned operators, of which two
have the quadratic divergence. At tree level, after EWSB, it generates
neutrino NSI and four-neutrino interactions\footnote{These are generally hard to
constrain directly and their effects are usually relatively weak, see, \eg,
\Refs~\cite{Bilenky:1999dn,Gavela:2008ra,Blennow:2008er}.} with strength
$\eps^{\delta \gamma}_{\beta \alpha}= -\frac{c^{\beta \alpha\delta
\gamma}}{2} \frac{v^4}{M^4}$. Notice that the
introduction of right-handed neutrinos already implies that NSI will
be induced at $d=6$ through the deviations from unitary
mixing. Consequently, the constraints derived from non-unitarity in
Ref.~\cite{Antusch:2008tz} would apply and the loop bounds would not
be so relevant. However, we find this toy example useful to connect
the quadratic divergence to a full theory in which it can be computed
and matched in order to clarify its interpretation. Indeed, since this
Standard Model extension is renormalizable and the corresponding
four-charged-fermion operator does not appear at tree level, all
diagrams are actually finite and we will be able to calculate them
unambiguously.  When computing the loops of the Higgs and Goldstones
in the full theory, we will assume that they are essentially massless
compared to the heavy mass scale $M$. We will also
assume that the external momenta are negligible. With these
approximations, the loop contribution is
\begin{eqnarray}
  \frac{c^{\alpha\beta\gamma\delta}}{16 \pi^2 M^2}
(\bar L_\alpha \gamma^\rho L_\beta)(\bar L_\gamma \gamma_\rho L_\delta)
  &=&\frac{c^{\alpha\beta\gamma\delta}}{16 \pi^2 M^2}\left[
  (\bar \nu_\alpha \gamma^\rho \nu_\beta)(\bar \nu_\gamma \gamma_\rho \nu_\delta) 
+ (\bar \ell_\alpha \gamma^\rho \ell_\beta)(\bar \ell_\gamma \gamma_\rho \ell_\delta)\right. \nonumber \\
&& \phantom{16 \pi^2 M^2}\left.
+ (\bar\nu_\alpha \gamma^\rho \nu_\beta)(\bar\ell_\gamma \gamma_\rho \ell_\delta)
+ (\bar\nu_\gamma \gamma^\rho \nu_\delta)(\bar\ell_\alpha \gamma_\rho \ell_\beta) \right]\, .
\label{1loop-ex}
\end{eqnarray}
It is easy to check that this is proportional to
$\mathcal{S}+\mathcal{A}$, as anticipated. We note that this operator
can be obtained from \eq~(\ref{eq:decomp}) simply by replacing
$|\phi_0|^2$ and $|\phi_+|^2$ by the factor of $M^2/(32
\pi^2)$ coming from the loop integral. Thus, the complete one-loop
four-fermion vertexes are given by replacing $|\phi_0|^2$ with
$v^2/2$, in order to determine the tree level contribution, and
replacing both $|\phi_0|^2$ and $|\phi_+|^2$ by
$M^2/(32 \pi^2)$ in order to determine the loop
contribution. From \eq~(\ref{1loop-ex}) we see that the following
interactions are generated, all with similar strength: four-neutrino
interactions, neutrino NSI, and four-charged-fermion
interactions. However, since the four-charged-fermion and the
four-neutrino interactions are completely symmetric under the exchange
of flavour indexes, while neutrino NSI are not, the
remaining terms for neutrino NSI and four-charged-lepton interactions are:
\begin{eqnarray}
\label{finaleps}
  \eps_{\alpha\beta}^{\gamma\delta} &=& 
  -\frac{v^2}{2M^2}\left[
    \frac{v^2}{M^2} (s^{\alpha\beta\gamma\delta}+a^{\alpha\beta\gamma\delta}
                     +x^{\alpha\beta\gamma\delta})
    +\frac{1}{8\pi^2}(s^{\alpha\beta\gamma\delta}+a^{\alpha\beta\gamma\delta})
    \right]\, ,\\
\label{finalepschf}
\eps^{\alpha\beta,\gamma\delta}_{CF,L} &=& 
-\frac{s^{\alpha\beta\gamma\delta}}{32\pi^2}\frac{v^2}{M^2}\, .
\end{eqnarray}
Here, $\eps^{\alpha\beta,\gamma\delta}_{CF,L}$ is defined through
\begin{equation}
  \mathcal L_{\rm CF} = -2\sqrt 2 G_F \eps^{\alpha\beta,\gamma\delta}_{CF,L}
  (\bar\ell_\alpha \gamma^\rho \ell_\beta)(\bar\ell_\gamma\gamma_\rho \ell_\delta)
\end{equation}
 Thus, if $8 \pi^2 v^2 \ll M^2$, both the
neutrino NSI and the four-charged-lepton operators will be dominated
by loop effects.

As we have just seen in the above example, a physical meaning can be
attributed to the quadratic divergences obtained in the effective
theory. Essentially, they can be regulated by reinserting the missing
propagators of the heavy particles in the full theory, leaving a
contribution to the effective NSI which is suppressed by
$v^2/(8\pi^2 M^2)$ instead of the tree level
$v^4/M^4$. In addition, the completely symmetric
contribution from $s^{\beta \alpha\delta \gamma}$ will generate an
additional four-charged-lepton operator at the one-loop level. Thus,
the $s^{\beta \alpha\delta \gamma}$ of a model could be severely
constrained by the strong bounds on decays such as $\mu \to 3e$, while the parameters $a^{\beta \alpha\delta \gamma}$ and
$x^{\beta \alpha\delta \gamma}$ only contribute to the neutrino NSI
and cannot be constrained from these processes. However, it is challenging to build a full theory which generates the antisymmetric couplings, but not the symmetric one, in a natural way.

An important caveat: in the above example of a full theory, we assumed
that the masses of the different heavy particles were the same. In
general, the tree and loop level contributions to the $\eps$s may be
different functions of the mass ratios and couplings, meaning that it could be
possible to fine-tune these functions in such a way that the
loop-contribution is zero, while still maintaining a non-zero
tree level contribution.


\section{Summary and Conclusions}

We have reconsidered the bounds on neutrino NSI from
one-loop processes. We have shown that, in order to have non-ambiguous
bounds, a gauge-invariant realisation of the NSI must be considered.
We explicitly studied $d=6$ and $d=8$ operators and have shown that,
in both cases, the logarithmic divergences of the one-loop
contributions are suppressed by the factor $m^2_\ell /M_W^2$, which
severely weakens the bounds.

In particular, for $d=6$ operators, the anti-symmetry relations that
arise as a consequence of the requirement of the absence of
four-charged-fermion interactions at tree level force the one-loop
processes to be zero in the absence of leptons masses.

In the $d=8$ case, we have shown that the loop processes involving NSI
should be calculated in the unitary gauge in order to obtain a gauge
invariant result if only the $W$ exchange
diagram is considered. In this way, the Goldstone loops present in gauge invariant
realisations of the NSI with extra Higgs doublets are automatically
taken into account. The
result is that the logarithmic divergence is proportional to the factor $m^2_\ell
/2M_W^2$. However, a quadratic divergence
is present and can be exploited to set bounds on NSI. The use of the
quadratic divergence in such a way implies model-dependent naturalness
assumptions, in particular on the coefficient of the divergence and
the size of the cut-off scale.

For a coefficient $k \simeq \mathcal O(1)$ and a cut-off $\Lambda$ of
the order of the electroweak scale, the bounds presented in
\Ref~\cite{Davidson:2003ha} are recovered. Pushing the cut-off scale
to $M$, where the effective operators are generated,
effectively assuming that no new physics appears between the
electroweak scale and $M$ to cancel the quadratic contribution,
implies that the loop processes can dominate over the tree level
contributions inducing four-charged-fermion interactions of a strength
similar to that of the neutrino NSI, unless $M \lesssim 1$~TeV. This allows the derivation of
strong bounds on the latter, but only based on naturalness arguments
and not model-independently. All these considerations apply to NSI of
neutrinos with both leptons and quarks. However, for neutrino NSI with
left-handed leptons, we have shown that, decomposing the NSI in parts
with different flavour symmetries, only the symmetric one contributes
to the four-charged-fermion process. Thus, if this part is not present
in the full theory from which NSI are realised, the loop constraints
can be avoided and large neutrino NSI are still viable if generated
via the antisymmetric couplings $a$ and $x$. It remains an open question 
whether there are natural models which can realise this. Otherwise an extra
fine-tuning beyond the one required at tree level would be necessary
to cancel the loop induced four-charged-fermion operators.

On the other hand, these naturalness arguments can always be evaded if
one allows fine-tuning of the theory and (given that
\Refs~\cite{Antusch:2008tz,Gavela:2008ra} have shown that large NSI
avoiding four-charged-fermion operators already require a significant
amount of fine-tuning at tree level) invoking naturalness arguments at
the loop level will not make the model more natural. Thus, we
emphasise the fact that no model-independent bounds can be derived
from the loop processes studied here, but that the only viable models
of large NSI avoiding four-charged-fermion interactions require
significant cancellations not only at tree level, but also at the one-loop level.

For practical purposes, in order to realise NSI without ad hoc
cancellations, the bounds derived in \Ref~\cite{Antusch:2008tz} have
to be respected or the neutrino NSI will be of the same order than the
four-charged fermion operators.  On the other hand, in a
model independent approach considering possible cancellations both at
tree and loop levels, the strongest effect is the loosening by three
orders of magnitude of the bound on $\eps_{e\mu}^{ff}$, since the
constraints in the other flavours are dominated by tree level
considerations \cite{Davidson:2003ha}. These tree level constraints of
$\mathcal O(10^{-1})$ still apply to $\eps_{e\mu}^{ff}$, but the
stringent radiative bounds of $\mathcal O(10^{-4})$ from $\mu \to 3e$ ($f=e$)
and $\mu \to e$ conversion in nuclei ($f=q$), is lost. Given the
strength of this bound, $\eps_{e\mu} = 0$ has been assumed for
simplicity in many phenomenological studies. Therefore,
it could be of interest to consider larger values for this parameter
in order to determine its impact on future neutrino oscillation
experiments. However, we would like to stress that this kind of large neutrino NSI would require significant fine-tunings at both the tree and loop levels.

\begin{acknowledgments}

We would like to thank S.~Antusch, S.~Davidson, A.~Donini, B.~Gavela, A.~Hoang, A.~Ibarra, V.~Mateu, M.~Papucci, C.~Pe\~{n}a-Garay,
S.~Pozzorini, G.~Raffelt, and I.~Stewart for
discussions and in particular N.~Rius and A.~Santamaria for very
useful and interesting discussions and for comments on this
manuscript. M.~B. and E.~F.-M. would also thank the people at IFIC and University of Valencia for their
kind hospitality during their visit.

This work was supported by the Swedish Research Council
(Vetenskapsr\aa{}det), contract no. 623-2007-8066 [M.~B.].

\end{acknowledgments}

\end{document}